# Extremely Nonperturbative Nonlinearities in GaAs Driven by Atomically Strong Terahertz Fields in Gold Metamaterials

C. Lange[1], T. Maag[1], M. Hohenleutner[1], S. Baierl[1], O. Schubert[1], E. Edwards[2], D. Bougeard[1], G. Woltersdorf[2], R. Huber[1]

[1]*Department of Physics, University of Regensburg, 93040 Regensburg, Germany*

[2]*Department of Physics, Martin-Luther-University Halle-Wittenberg, 06120 Halle, Germany*

**Terahertz near fields of gold metamaterials resonant at a frequency of 0.88 THz allow us to enter an extreme limit of non-perturbative ultrafast THz electronics: Fields reaching a ponderomotive energy in the keV range are exploited to drive nondestructive, quasi-static interband tunneling and impact ionization in undoped bulk GaAs, injecting electron-hole plasmas with densities in excess of $10^{19}$ cm$^{-3}$. This process causes bright luminescence at energies up to 0.5 eV above the band gap and induces a complete switch-off of the metamaterial resonance accompanied by self-amplitude modulation of transmitted few-cycle THz transients. Our results pave the way towards highly nonlinear THz optics and optoelectronic nanocircuitry with sub-picosecond switching times.**

Intense, phase-locked light pulses in the terahertz (THz) spectral range have opened up an exciting arena for field-sensitive nonlinear optics [1-22]. For a given peak electric field $\mathcal{E}$, the low carrier frequency $\omega_{THz}$ gives rise to a potentially large ponderomotive energy $U_p = e^2\mathcal{E}^2/4m\omega_{THz}^2$, which quantifies the cycle-averaged quiver energy of a free electron of mass $m$. This situation promises a new quality of non-perturbative light-matter interaction at the boundary of THz optics and high-speed electronics.

For frequencies between 0.5 and 3 THz, optical rectification [7-10] has enabled transients with peak field amplitudes in excess of 1 MV/cm [10]. Using such an electromagnetic pulse as an alternating bias, THz-driven carrier-multiplication in doped semiconductors [12] and graphene [14] has been demonstrated. Strong THz fields have also been used to drive spectacular nonlinear intraband dynamics of quasiparticles, such as field ionization of impurity states [13] or excitons, electron-hole recollisions and high-order sideband generation [11]. In these experiments, the THz bias facilitates tunneling of bound electrons through potential energy barriers, which correspond to binding energies between a few meV and several 10 meV [Fig. 1(a)] [11,13,23].

A new limit of non-perturbative nonlinearities is expected if THz amplitudes approach atomically strong fields. As depicted in Fig. 1(b), for typical semiconductors like GaAs, hypothetical THz amplitudes of 12 MV/cm would cause a transient potential energy drop by the fundamental energy gap $E_g = 1.4$ eV over distances as small as two unit cells, and field-induced interband tunneling should start to dominate. Using a band model [24], we estimate that under such conditions more than $10^{20}$ valence electrons per cm$^3$ may undergo Zener tunneling through the fundamental bandgap into the conduction band, within a time window of only 100 fs, possibly inducing even degenerate carrier populations in intrinsic bulk semiconductors [25].

Phase-locked light pulses featuring amplitudes in the 100 MV/cm range have become available at carrier frequencies of multiple 10 THz [3]. This breakthrough has enabled ultrafast biasing of bulk solids in an unprecedented high-field limit, where a coherent interplay between non-resonant interband polarization and intraband Bloch oscillations generates THz high-harmonic radiation [6]. The diffraction limit of focusing, however, has precluded comparably high fields at frequencies as low as 1 THz where yet larger ponderomotive potentials $U_p$ combined with photon energies orders of magnitude below electronic interband resonances could pave the way to quasi-static biasing. Custom-tailored metamaterials are a promising concept for overcoming the diffraction limit. Indeed, field enhancement in metamaterials has been exploited to induce a metal-insulator phase transition in VO$_2$ by a THz transient with a carrier frequency below 1 THz [26], and to evoke strong THz nonlinearities through THz-driven intervalley scattering and impact ionization [22].

Here, we combine metamaterials with MV/cm-scale few-cycle sub-1-THz transients to explore an extreme regime of non-perturbative nonlinearities in the well-defined laboratory of the bulk semiconductor GaAs. Atomically strong fields exceeding 10 MV/cm and ponderomotive potentials in the keV range lead to strong carrier generation across the band gap which exceeds the central THz photon energy by a factor of 390. Ultrabroadband near-infrared and visible interband luminescence emerges, and a strong THz nonlinearity results on the single-cycle time scale.

Near-infrared light pulses centered at a wavelength of 800 nm (pulse energy: 3.5 mJ, pulse duration: 35 fs) are derived from a Ti:sapphire laser amplifier. We employ tilted-pulse front optical rectification [9] in a cryogenically cooled LiNbO$_3$ crystal to generate phase-locked THz transients [Fig. 1(c)] with a peak field of $\mathcal{E}_0 = 1.5$ MV/cm in the far-field focus on the sample, and an amplitude spectrum ranging from 0.1 to 2.5 THz [Fig. 1(d)]. For electro-optic detection of the THz waveform, the transmitted beam is imaged onto a 10-µm-thick ZnTe detection crystal and gated by 35-fs near-infrared pulses.

The unit cell of our metamaterials is given by a planar gold resonator [Fig. 1(e)] featuring a 2.5-µm-wide double capacitive gap, symmetrically closed by inductive loops on each side. This miniature RLC circuit of outer dimensions of 37.5 µm × 30 µm is designed for a resonance frequency of 0.88 THz. Millimeter-sized, two-dimensional arrays of identical resonators made of 150-nm-thick gold are fabricated by electron-beam lithography, thermal evaporation and lift-off processes. The gold structure and a 3-nm-thick chromium adhesion layer are created either directly on a substrate of undoped (100)-oriented GaAs (sample A) or on top of an additional Al$_2$O$_3$ layer of a thickness of $t = 100$ nm (sample B) or $t = 300$ nm (sample C). Figure 1(e) shows the typical distribution of the electric near-field in the unit cell of the metamaterial, on resonance, as computed by finite-difference frequency-domain (FDFD) calculations. In the capacitive gap region of the structure, the electric field is strongly enhanced, whereas the magnetic field enhancement is negligible (not shown) and thus not further considered.

Near the capacitive stripes of our metamaterials, time-domain simulations predict an enhancement factor between the near- ($\mathcal{E}_n$) and the far-field ($\mathcal{E}_0$) of up to 35 for our THz pulse shape [25]. For the highest

THz far fields attained in our experiment this enhancement would correspond to $\mathcal{E}_n > 50$ MV/cm if no nonlinearities occurred. To date, there is no calibrated procedure to characterize such enormously large fields. Established near-field measurements relying on electro-optic sampling [27, 28], tunneling currents [29] or Franz-Keldysh oscillations [30] are not characterized for atomically strong fields in which even the concept of a rigid electronic bandstructure is pushed to the limits of its validity. Possible characterization methods may, however, arise from the following observations.

We illuminate our structure with intense THz pulses and map the metamaterial plane onto a silicon CCD camera or a grating spectrometer using a microscope objective. The strong near-field enhancement in the metamaterial manifests itself in a spectacular way: Resonators located in the THz focus ($\mathcal{E}_0 = 1.5$ MV/cm) exhibit bright visible electroluminescence (EL) in GaAs. False-color microscope images of the spectrally integrated luminescence intensity emanating from a selected area of metamaterial B and close-up views of individual THz resonators in all three samples are shown in Figs. 2(a-d), respectively. For a direct comparison, a map of the calculated local field enhancement is shown in Fig. 2(e). The spatial distribution of the emitted radiation correlates strongly with the local field: Most luminescence originates from the region close to the stripes of the capacitive double-gap area, where the field enhancement is largest, while emission is virtually absent in unstructured areas of the substrate. Typical spatial profiles of the EL intensity [cf. Fig 2(f)] exhibit features of a full-width at half-maximum of less than 950 nm and allow for indirect tracing of the local THz field amplitude.

The microscopic origin of the strong light emission is apparent from its spectral shape. Figure 3(a) shows typical EL intensity spectra obtained for a series of THz amplitudes. For $\mathcal{E}_0 = 120$ and $380$ kV/cm, the luminescence extends from the direct band gap of GaAs at $E_g = 1.42$ eV up to 1.7 eV, attesting to the presence of a dense and hot electron-hole plasma in the substrate [31]. At amplitudes above 800 kV/cm, luminescence at yet higher photon energies of up to $\hbar\omega = 1.9$ eV is observed. Additionally, a shoulder develops in the spectra at 1.65 eV (black arrow). For an energy 50 meV above the band gap, $I_L$ scales as $|\mathcal{E}_0|^{1.1}$. Still at high excess energies of 300 meV a low-order power law of $I_L \sim |\mathcal{E}_0|^{1.5}$ is found [Fig. 3(b)], characteristic of non-perturbative carrier generation [6].

For an estimate of the density of field-induced electron-hole pairs, we model the interband luminescence as a three-step process: (i) carrier generation within the duration of the THz transient, (ii) thermalization and cooling of the electron-hole plasma, and (iii) interband recombination. In high-purity, direct-gap semiconductors, the recombination time greatly exceeds the duration of carrier generation and thermalization. Therefore, we describe the electron-hole plasma by a hot carrier distribution and estimate an average carrier temperature $T$ and density $n$ by fitting the spectra using a convolution of a Fermi-Dirac distribution with the joint density of states in the effective mass approximation (Kubo-Martin-Schwinger relation). This procedure works reasonably well up to $\mathcal{E}_0 = 380$ kV/cm, where we obtain $T \approx 1200$ K and a sizable carrier concentration of $n = 5 \times 10^{18}$ cm$^{-3}$. However, it cannot reproduce spectra featuring a prominent shoulder at 1.65 eV, as observed for higher field amplitudes. For large carrier excess energies, the effective mass approximation and the assumption of momentum-independent dipole moments falter and scattering into the L-valleys of GaAs sets in. Therefore, at $\mathcal{E}_0 = 1.5$ MV/cm, we restrict ourselves to fitting the EL spectrum in the energy range below 1.65 eV [black dotted curve in Fig. 3(a)]. Best agreement with the data is obtained for an average electron temperature of $T \approx 2000$ K, a carrier concentration as high as $n = 1.5 \times 10^{19}$ cm$^{-3}$, and an electron filling factor at the Γ point of 0.8, indicative of a degenerate carrier population. Hence, one may already attain population inversion in the most highly excited areas. Note that the above densities represent lower bounds of the actual peak values since the luminescence spectra are the result of spatial integration.

In order to understand the microscopic mechanism of THz-driven carrier generation in our structure, we refine our estimate of the near-field THz amplitude including near-field screening by the electron-hole plasma with $n = 1.5 \times 10^{19}$ cm$^{-3}$. To this end, we attribute a Drude conductivity to those sub-volumes of the substrate which exhibit the highest near-field enhancement [25]. Although these simulations do show a reduction of the near-field amplitude, they still predict a maximum $\mathcal{E}_n > 20$ MV/cm [25]. The calculation underestimates the actual field since it assumes a time-independent carrier density that is already present at the leading edge of the THz pulse and neglects the dynamical buildup of screening [32]. We base our following considerations on a very conservatively chosen near-field amplitude of $\mathcal{E}_n = 12$ MV/cm for the intuitive picture sketched in Fig. 1(b), noting however that the actual fields are even larger.

In contrast to previous studies exploring field-driven EL in doped semiconductors [13,23], our substrates are undoped. The initial pair generation, therefore, has to start from THz-induced interband transitions. Whether multi-photon ionization or interband tunneling prevails can be determined via the Keldysh parameter $\gamma_K = \omega_{\mathrm{THz}}\sqrt{2mE_g}/e\mathcal{E}_n$ [33]. While the quantum nature of light dominates for $\gamma_K \gg 1$, tunneling injection becomes efficient for $\gamma_K \lesssim 1$. Strong THz-driven interband transitions in the regime of $\gamma_K \approx 0.13$ were recently demonstrated utilizing multi-MV/cm fields at 30 THz, and confirmed through a microscopic theory of interband polarization and intraband currents [6]. The electric peak field $\mathcal{E}_n$ in our present experiment approaches similar values as in Ref. 6, while $\hbar\omega_{THz} = E_g/390$ is substantially smaller. Therefore we arrive at a record-low $\gamma_K \approx 2 \times 10^{-2}$, fostering quasi-static Zener tunneling from the valence- to the conduction band as an efficient first step of carrier generation. According to the rough estimate given above [Fig. 1(b)], Zener tunneling alone may already account for the carrier densities observed in the experiment [34]. Yet, we point out that our estimate is based on a bandstructure model [24] which is pushed to its limits in atomically strong fields.

In the absence of scattering, the THz bias would drive the wavevectors $k$ of Zener-injected carriers through the full Brillouin zone according to Bloch's acceleration theorem, ultimately giving rise to coherent Bloch oscillations [6]. We do not expect this scenario for the current situation since the wide conduction band of GaAs allows for efficient impact ionization once the carrier excess energy exceeds the threshold energy $E_{th} = E_g(2m_e+m_{hh})/(m_e+m_{hh}) = 1.6$ eV, where $m_e$ and $m_{hh}$ are the electron and heavy hole masses [13,24]. For $\mathcal{E}_n \approx 12$ MV/cm, electrons require less than 10 fs to gain sufficient energy for a single impact ionization event. One half-cycle of the THz transient should thus drive a cascade of at least 10 ionization steps, multiplying the carrier density by $2^{10} \approx 10^3$ per oscillation half cycle. The same multiplication factor can also be estimated considering that the ponderomotive energy $U_p = e^2\mathcal{E}_n^2/4m\omega_0^2 \approx 2$ keV $\approx 10^3 \times E_{th}$, i.e., sufficient to generate $10^3$ electron-hole pairs. Note that this multiplication effect can only fully apply as long as Pauli blocking does not limit this process. In fact, such saturation effects due to degenerate carrier populations should play an important role in our experiment and may account for the scaling of $I_L$ with a low power of $\mathcal{E}_0$ [Fig. 3(b)]. The regime explored here is in qualitative contrast to previous experiments where field-ionization of shallow impurities by the far-field of strong THz pulses has been used to observe luminescence with a field-independent narrow bandwidth ($\approx 5$ meV), $I_L \sim \mathcal{E}_0^8$ and carrier densities between $10^{13}$ cm$^{-3}$ and $10^{16}$ cm$^{-3}$ [13].

Besides the local field enhancement, metamaterials offer a second functionality: They couple the THz response of the microscopic volume hosting the dense electron-hole plasma back into the far field [22,26]. Figure 4 demonstrates that the extremely sub-wavelength interaction regions of the sample induce a complete suppression of the metamaterial resonance, causing a self-amplitude modulation of the THz pulses. The transmitted waveform for $\mathcal{E}_0 = 150$ kV/cm [Fig. 4a] exhibits trailing oscillations characteristic of the resonant metamaterial response for delay times t > 2 ps. These features vanish completely if the THz peak amplitude is increased to $\mathcal{E}_0 = 1.5$ MV/cm [Fig. 4b]. In the frequency domain this switch-off of the resonance manifests itself in a change of the relative field transmission from 50% ($\mathcal{E}_0 = 150$ kV/cm) to nearly unity ($\mathcal{E}_0 = 1.5$ MV/cm), at a frequency of 0.88 THz [Fig. 4(c)]. Simultaneously, the metamaterial resonance shifts from 0.88 THz to 0.75 THz.

The transients also illustrate the time frame on which the strong THz nonlinearity establishes itself. Transmitted waveforms follow the same time dependence as the incident transient up to a delay time of $t = 0$ fs [Figs. 4(a) and (b)]. At the subsequent three field extrema, the transmitted amplitude is steadily reduced in the high-field limit [Fig. 4(b)] due to accumulative carrier generation occurring at a progressively increasing rate in each half-cycle of the THz pulse. We therefore expect that a significant reduction of the near-field amplitude by the dense electron-hole plasma occurs only for the later half-cycles where the majority of carriers are generated.

Using FDFD simulations we model the far-field THz response caused by carrier injection, assuming conductive slabs in the substrate below the capacitive gap region [25]. Calculated spectra are shown in Fig. 4(d) for carrier densities of $n = 10^{16}$ to $10^{19}$ cm$^{-3}$. Since our simulation neither accounts for the temporal dynamics nor for the precise spatial density profile of the carrier population, the densities used here mark a lower bound for the actual peak values. Despite these simplifications the nonlinear response of the metamaterial is reproduced qualitatively well: With increasing carrier density, the RLC circuit's loss rate rises and the oscillation is strongly damped, resulting in a reduction of the resonance frequency and a decrease of the oscillator strength.

In conclusion, we have demonstrated a new class of extremely non-perturbative nonlinearities in a semiconductor with an unprecedented ponderomotive potential reaching the keV range and a Keldysh parameter of only $\gamma_K \approx 2 \times 10^{-2}$. The strong fields allow us to inject degenerate electron-hole populations, leading to broadband interband electroluminescence at energies more than 500 times the THz photon energy, and nonlinear THz transmission through the sample. With increasing field amplitude the resonance of the metamaterial sample is red-shifted and finally switched off completely. Exploiting non-destructive near-field enhanced THz-biasing we reach record field amplitudes of at least 12 MV/cm for center frequencies below 1 THz. This results is an important step towards high-field studies for electronics at THz clock rates. For instance, the perspective to reach population inversion in custom-cut nanoscale regions makes our metamaterials an interesting platform for sub-micron optoelectronics at THz speeds. The strongly nonlinear THz response allows for novel devices such as THz transistors with femtosecond precision, spectrally selective switches and low-threshold saturable absorbers, all flexible in frequency and bandwidth by the choice of metamaterial.

This work was supported by the European Research Council through ERC grant 305003 (QUANTUMsubCYCLE), and the Deutsche Forschungsgemeinschaft through projects LA 3307 and collaborative research center SFB 689.

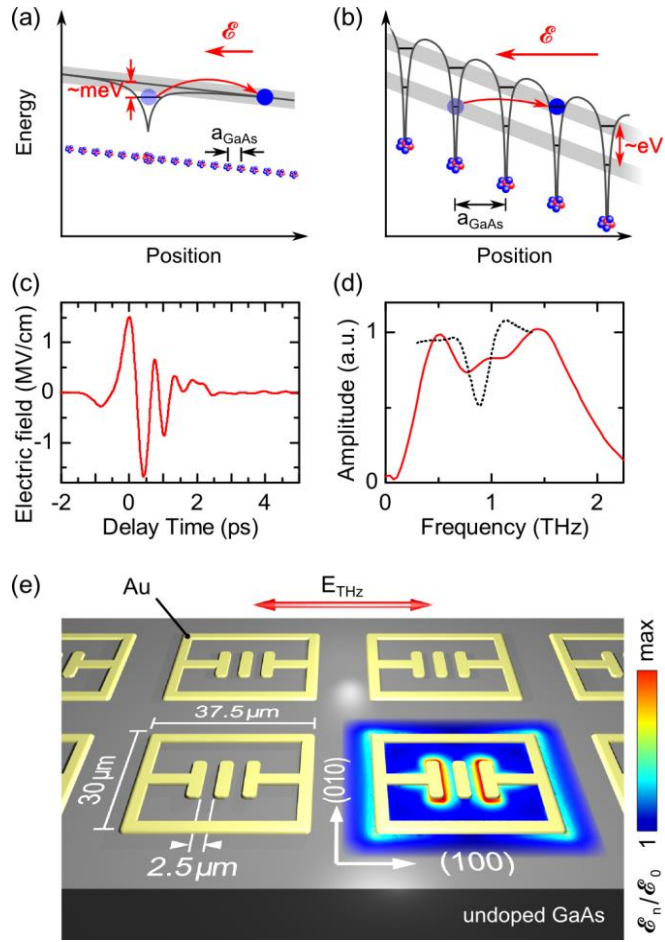

Fig. 1: (a) A THz field of $\mathcal{E} = 500$ kV/cm skews the conduction band edge (solid line) of GaAs (lattice constant $a_{GaAs}$) sufficiently to drive field ionization of shallow impurity states (blue sphere). (b) Under an atomically strong bias of $\mathcal{E} = 12$ MV/cm the energy landscape is so strongly tilted that valence and conduction band-like states of next-neighboring unit cells become energetically degenerate and massive interband Zener tunneling should set in. (c) Far-field waveform of phase-locked few-cycle THz transients (amplitude: 1.5 MV/cm), (d) spectrum (red line), and normalized metamaterial transmission (dotted curve). (e) Geometry of the resonator forming the unit cell of the THz metamaterial on (100)-oriented GaAs as indicated. The red double arrow marks the THz polarization for optimal coupling to the fundamental resonator mode. The near-field enhancement of the structure at the resonance frequency is highlighted as a color plot for the lower right resonator. This map has been calculated by FDFD simulation for a plane 300 nm below the surface of the substrate of sample A.

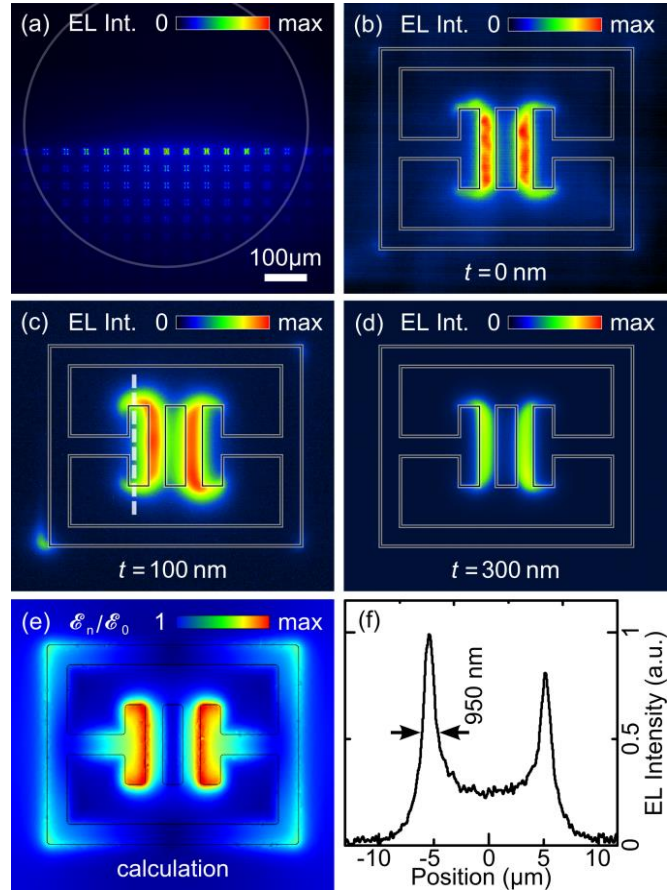

Fig. 2: (a) THz-induced EL from the metamaterial array. The THz focus (full-width at half-maximum outlined by white circle) is positioned at the horizontal border between the metamaterial field (lower half) and the unstructured GaAs substrate (upper half), where no emission is observed. (b) Microscopic view of the EL intensity from a single resonator (sample A, thickness of the insulating layer $t = 0$ nm) driven by the THz transient of Fig. 1(c) ($\mathcal{E}_0 = 1.5$ MV/cm). The outline of the resonator is superimposed on the luminescence image as a guide to the eye. (c),(d) EL intensity images from resonators of samples B and C, with $t = 100$ nm and $t = 300$ nm, respectively. (e) Color map of calculated near-field distribution indicating the modulus of the electric field amplitude at the resonance frequency of 0.88 THz, in a plane 300 nm below the semiconductor surface. (f) The EL intensity profile across the capacitor plate (dotted line in (c)) demonstrates a spatial resolution better than 950 nm.

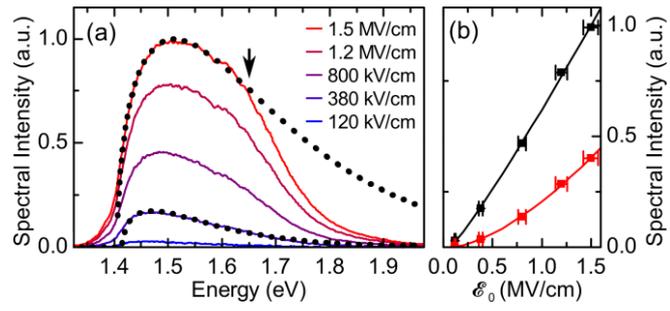

Fig. 3: (a) Luminescence spectra collected from sample B for different field amplitudes $\mathcal{E}_0$ (color-coded, solid curves). Black dotted curves: numerical fit functions (see text for details). (b) Scaling of EL intensity at photon energies 50 meV (black) and 300 meV (red) above the fundamental bandgap with the THz field (squares). Solid curves: Power-law fit functions (see text).

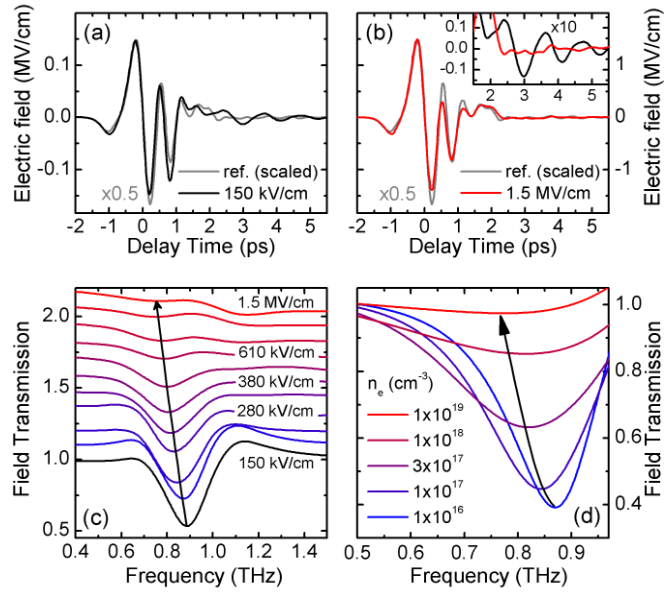

Fig. 4: Nonlinear THz transmission through the metamaterial sample. (a) Transmitted THz waveform (black curve) for a peak field of $\mathcal{E}_0 = 150$ kV/cm. (b) Corresponding transient for $\mathcal{E}_0 = 1.5$ MV/cm (red curve). Gray curve: scaled waveform of incident THz transient. Inset: magnified view of waveforms in (a) and (b). (c) Transmission spectra for different THz peak fields, vertically offset for clarity. For $\mathcal{E}_0 = 150$ kV/cm (black curve) a sharp transmission minimum at 0.88 THz is observed, which red-shifts (black arrow) and bleaches for increasing $\mathcal{E}_0$. (d) Numerical simulation of the spectral response of the structure for a series of carrier concentrations in the gap region using an electron mobility in GaAs of $\mu_{GaAs} = 2600$ cm/Vs. Black arrows are guides to the eye.